\documentclass{article}
\usepackage{ijcai24}

\usepackage{times}
\usepackage{soul}
\usepackage{url}
\usepackage[hidelinks]{hyperref}
\usepackage[utf8]{inputenc}
\usepackage[small]{caption}
\usepackage{graphicx}
\usepackage{amsmath}
\usepackage{amsthm}
\usepackage{booktabs}
\usepackage[switch]{lineno}
\usepackage{subfig}
\usepackage[ruled,linesnumbered]{algorithm2e}
\usepackage{multirow}
\usepackage[normalem]{ulem}
\useunder{\uline}{\ul}{}
\usepackage{bm}
\usepackage{setspace}
\usepackage{enumitem}
\usepackage{amsfonts}
\usepackage {color}

\title{Dynamicity-aware Social Bot Detection with Dynamic Graph Transformers}
\author{
Buyun He$^{1,}$\footnotemark[1] \and
Yingguang Yang$^{1,}$\footnotemark[1]\and
Qi Wu$^{1}$ \and
Hao Liu$^1$\and
Renyu Yang$^2$ \\
Hao Peng$^{2,3,}$\footnotemark[2]\and
Xiang Wang$^1$\and
Yong Liao$^{1,}$\footnotemark[2]\and
Pengyuan Zhou$^1$
\\
\affiliations
$^{1}$University of Science and Technology of China, $^{2}$Beihang University,\\ $^{3}$Harbin Engineering University\\
\emails
\{byhe, dao, qiwu4512, rcdchao\}@mail.ustc.edu.cn,
\{renyu.yang, penghao\}@buaa.edu.cn,
xiangwang1223@gmail.com,
\{yliao, pyzhou\}@ustc.edu.cn
}

\begin{document}
\maketitle
\begin{abstract}
  Detecting social bots has evolved into a pivotal yet intricate task, aimed at combating the dissemination of misinformation and preserving the authenticity of online interactions. While earlier graph-based approaches, which leverage topological structure of social networks, yielded notable outcomes, they overlooked the inherent dynamicity of social networks -- In reality, they largely depicted the social network as a static graph and solely relied on its most recent state. Due to the absence of dynamicity modeling, such approaches are vulnerable to evasion, particularly when advanced social bots interact with other users to camouflage identities and escape detection. To tackle these challenges, we propose BotDGT, a novel framework that not only considers the topological structure, but also effectively incorporates dynamic nature of social network. Specifically, we characterize a social network as a dynamic graph. A structural module is employed to acquire topological information from each historical snapshot. Additionally, a temporal module is proposed to integrate historical context and model the evolving behavior patterns exhibited by social bots and legitimate users. Experimental results demonstrate the superiority of BotDGT against the leading methods that neglected the dynamic nature of social networks in terms of accuracy, recall, and F1-score. %% Further studies bear out the effectiveness of incorporating the dynamic nature of social networks for social bot detection.

\end{abstract}

\maketitle
\renewcommand{\thefootnote}{\fnsymbol{footnote}} 
\footnotetext[1]{The authors contributed equally to this work.}
\footnotetext[2]{Corresponding authors.}
\vspace{-1em}
\section{Introduction}
As social networks become integrated into people's daily routines, there is a prevalent occurrence of program-controlled bots masquerading as legitimate users for malicious purposes~\cite{subrahmanian2016darpa}. Social bots engage in detrimental activities such as propagating misinformation~\cite{varol2017online,gao2023rumor}, manipulating public opinion~\cite{cui2020deterrent}, interfering in elections~\cite{rossi2020detecting} and promoting extremist ideologies~\cite{ferrara2016predicting}. It is therefore imperative to effectively detect social bots to mitigate the detrimental societal and economic impact and to preserve the integrity of social network information.

\begin{figure}
  \centering
\includegraphics[width=0.9\linewidth]{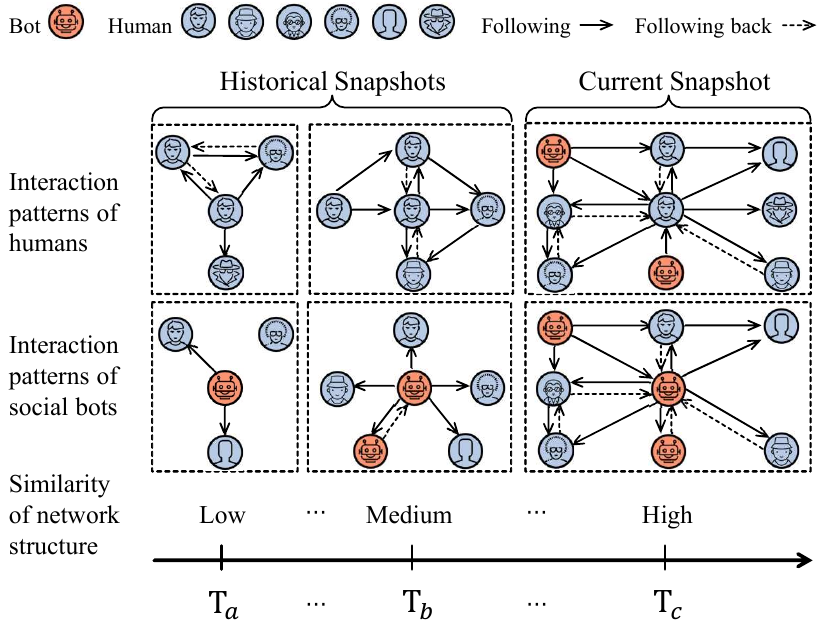}
  \vspace{-0.5em}
  \caption{Dynamic nature of social network}
  \label{fig:intro}
  \vspace{-1.75em}
\end{figure}

Traditional techniques for bot detection are largely based on features, requiring extraction of either numerical feature from user information~\cite{yang2013empirical} or semantic features from textual information~\cite{wei2019twitter,dukic2020you}. However, bot operators can often bypass bot detection through advanced countermeasures, which is commonly referred to as \textit{bot evolution}~\cite{cresci2020decade}. In fact, the detectability of the feature-based methods is vulnerable to imitation and evasion, as bot operators can effortlessly steal user information from legitimate users or intersperse a few malicious messages with many neutral ones~\cite{feng2022twibot}. As a result, such methods are inaccurate in spotting disguised social bots. With the advancements in graph neural networks, some researchers employed graph-based methods~\cite{wu2023heterophily,feng2022heterogeneity,yang2023rosgas} to identify the disguised social bots. They typically assume that the network structure of social bots generally differs from that of legitimate users. For instance, social bots tend to have sparser connections and randomly select users to interact with, whereas human beings prefer to connect with others who share similar characteristics~\cite{yang2013empirical}. These graph-based methods are among top performers by leveraging the topological structure of social networks for bot detection. However, most of the existing graph-based detection methods interpret the social network as a \textit{static} graph and fail to acquire the dynamic nature of social networks. As shown in Figure \ref{fig:intro},  there still remain two intractable issues:

\textbf{Deficiency in utilizing historical interaction graph context.} Similar to the case of evading detection from feature-based methods by forging numerical or semantic features, the ever-evolving social bots are meticulously engineered to interact with legitimate users and mimic their network structures~\cite{cresci2020decade} to escape graph-based detection. However, despite the structure of social network has changed, the discrepancies in the previous interaction graph between social bots and benign users could reveal the deception of social bots and uncover their true identity. Unfortunately, conventional approaches upon \textit{static} graphs solely rely on the last state of the social network and overlook the valuable historical interaction graph context. Consequently, if the social bots have \textit{already} completed their disguise, it is challenging for static graph based methods to distinguish benign users from the evolved social bots.

\textbf{Limitation of modeling evolving behavior patterns.} Social bots evolve over time, evading detection by dynamically adapting their actions, strategies, or interaction patterns to mimic legitimate users. In contrast, genuine users do not require such adaptations and exhibit different evolution of behavior patterns compared to social bots. Discovering the evolving behavior patterns may enhance the effectiveness of social network modeling~\cite{liu2020detecting}. Nevertheless, static graph based methods fall short of modeling the distinct evolving behavior patterns of social bots and legitimate users, leading to erroneous results when conducting bot detection tasks. 

%%In summary, a real-world social network is inherently \textit{dynamic}, formed through temporal interactions among users~\cite{aouay2014modeling}. However, the dynamic nature of social networks remains largely unexplored in graph-based bot detection methods, which leads to the deficiency in modeling past interaction graph context and the limitation of modeling the distinct evolution of behavior patterns exhibited by social bots and legitimate users. 

To overcome the limitations above, we propose a new framework called BotDGT (\textbf{Bot} detection with \textbf{D}ynamic \textbf{G}raph \textbf{T}ransformers). The key insight is to introduce the dynamicity modeling of social network for bot detection. To this end, BotDGT depicts a social network as a \textit{dynamic} graph for modeling historical interaction graph contexts and discerning the evolving behavior patterns. Specifically, we interpret users and interactions as nodes and edges, respectively, to generate a batch of snapshots at a fixed time interval for a given social network. A structural module that employs message-passing mechanism is proposed to model the topological structure of each historical snapshot. Additionally, a temporal module based on self-attention mechanism is further employed to incorporate historical contexts and exploit the distinct behavior patterns evolution exhibited by social bots and legitimate users. Overall, our contributions are summarized as follows:
\begin{itemize}[leftmargin=*]
  \item To the best of our knowledge, we are the first to characterize a social network as a dynamic graph and effectively identify the ever-evolving social bots that disguise themselves through adapting their behavior patterns.
  \item We introduce a novel bot detection framework to consider both topological structure and the dynamic nature of social networks to enhance the performance of bot detection.
  \item We conduct comprehensive experiments on two benchmarks for bot detection, which demonstrates the superior performance of BotDGT compared to the leading methods in terms of accuracy, recall and F1-score. Further experiments substantiate the effectiveness of incorporating the dynamic nature of social networks for bot detection.
\end{itemize}
% \vspace{-1em}
\section{Preliminaries}

\subsection{Related Work}
\subsubsection{Social Bot Detection}
Early methods for social bot detection are predominantly feature-based. Researchers extracted numerical features from user information and fed them into machine learning models for classification~\cite{lee2011seven,mazza2019rtbust,yang2020scalable} or anomaly detection~\cite{miller2014twitter}. Some studies employed natural language processing techniques to encode textual information, capturing semantic features to enhance the feature-based methods~\cite{hayawi2022deeprobot,dukic2020you,yang2023fedack}. However, newer generations of social bots may forge numerical features or semantic features, either by stealing legitimate users’ information or interspersing malicious messages among benign ones, to evade feature-based detection.

With the advancements in graph neural networks, some graph-based methods leveraged the topological structure of social networks for bot detection~\cite{pham2022bot2vec,shi2023over,peng2024unsupervised,zeng2024adversarial}. The study~\cite{ali2019detect} takes the first attempt to introduce graph convolutional networks to aggregate user information from neighboring nodes for bot detection. Subsequent investigations modeled the heterogeneity of social networks and yielded leading performance~\cite{feng2021botrgcn,feng2022heterogeneity}. However, these methods interpreted the social networks as static graphs and neglected the intrinsic dynamicity of real-world social networks, thereby falling short of detecting evolving social bots that adapt strategies to mimic legitimate users' network structure~\cite{cresci2020decade}. To this end, we build upon previous research and present a dynamicity-aware bot detection framework. It incorporates historical context and exploits the evolution of user behavioral patterns, aiming at enhancing the performance of bot detection.

\begin{figure*}[t]
    \centering
    \includegraphics[width=0.858\linewidth, height=0.3756\linewidth]{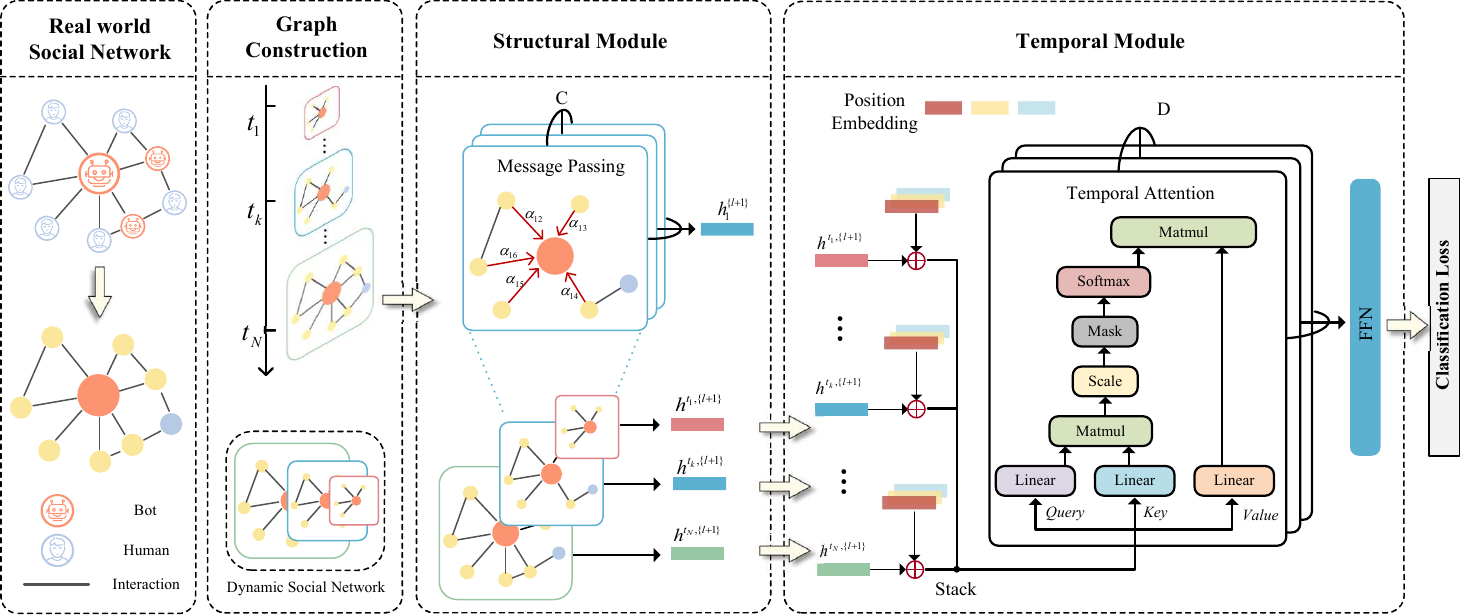}
    \caption{Overview of our proposed BotDGT framework.}
    \vspace{-1.5em}
    \label{fig:framework}
\end{figure*}

\subsubsection{Dynamic Graph Neural Network}
Dynamic graphs capture temporal information through time-based dimensions~\cite{skarding2021foundations,peng2021lime}. Previous research in graph representation learning has predominantly concentrated on static scenarios, presuming fixed topological structures. However, real-world graphs, including social networks~\cite{alvarez2021evolutionary,wang2021tedic}, exhibit continual evolution and dynamic characteristics over time. Dynamic graph neural networks are designed to capture this dynamic nature and are widely adopted in various tasks, including link prediction~\cite{xie2021learning,chen2022gc,sankar2020dysat,zhang2023attentional}, anomaly detection~\cite{cai2021structural,guo2022subset}, and node classification~\cite{kim2022dygrain,pareja2020evolvegcn,xu2020inductive}. Drawing upon the previous works that employed recurrent neural networks~\cite{chen2022gc,zhou2020heterogeneous} and attention mechanisms~\cite{sankar2020dysat,xu2020inductive} to model dynamic graphs, we propose a novel approach that utilizes a self-attention mechanism to leverage the dynamic nature of social network, thereby enhancing the effectiveness of bot detection.

\subsection{Problem Definition}
In this paper, we depict the social network as a dynamic graph that changes over time and capture the dynamic nature of the social network to improve the performance of the bot detection model. In this part, we first define the dynamic social network and then formulate the problem.
% \newtheorem*{def1}{Definition 1 (social network graph)} 
% \begin{def1}
% A social network graph is defined as a graph $G = (V, E)$ where $V$ is the set of nodes, $E$ is the set of edges, $N$ is the set of node types and $R$ is the type of edge types. Each node $v\in V$ and each edge $e\in E$ are associated with their type mapping functions $\varphi_{1}:V\to N$ and $\varphi_{2}:E\to R$, respectively.
% \end{def1}
\newtheorem*{def2}{Definition} 
\begin{def2}[Dynamic Social Network]
A dynamic social network is depicted as a graph $G =\left \{ G^{t_{1}}, G^{t_{2}},..., G^{t_{N}}\right \} $ with a series of network snapshots over time. $G^{t_{k}}=(V^{t_{k}}, E^{t_{k}})$ represents the snapshot of a given social network graph at the timestamp $t_{k}$, where $V^{t_{k}}$, $E^{t_{k}}$ are users and interactions respectively observed at timestamp $t_{k}$.
\end{def2}
% \newtheorem*{def3}{Definition 3 (dynamic social network representation learning)} 
% \begin{def3}
% Given a dynamic social network, $G =\left \{ G^{t_{1}}, G^{t_{2}},..., G^{t_{N}}\right \} $, dynamic social network representation learning aims to learn a mapping function $f: v\to \mathbb R^{d}$ for each node $v \in V^{t_{k}}$ at the timestamp $t_{k} \in \left \{t_{1},t_{2},...,t_{N}  \right \} $, where $d$ is the number of embedding dimensions. The mapping function $f$ should capture both the topological structure and the dynamic nature of the social network.
% \end{def3}
To align with the previous studies, we treat bot detection as a binary classification problem, i.e., users are classified into human ($y$ = 0) or bot ($y$ = 1). We formulate the problem of bot detection in dynamic social networks as follows:
\newtheorem*{def4}{Problem} 
\begin{def4}[Bot Detection in Dynamic Social Network]
Given a dynamic social network $G =\left \{ G^{t_{1}}, G^{t_{2}},..., G^{t_{N}}\right \} $, the problem is to find an encoding function $f : v\to \hat{y} $ for each node $v \in V^{t_{k}}$ at the timestamp $t_{k} \in \left \{t_{1},t_{2},...,t_{N}  \right \} $, such that $\hat{y}$ approximates the 
ground truth $y$ to maximize prediction accuracy.
\end{def4}
% \vspace{-0.5em}
\section{Methodology}
As shown in Figure \ref{fig:framework}, BotDGT comprises two modules to primarily capture the topological structure and dynamicity of social networks. Specifically, we produce snapshots of the social network at a certain time interval. For each snapshot, the structural module aggregates the features of neighboring nodes and generates temporary node representations containing the snapshot's topology information. Additionally, the temporal module integrates historical context and exploits changes in behavior patterns over time. The resultant node representations are subsequently utilized to differentiate social bots from genuine users.

\subsection{Constructing a Dynamic Graph}
We first construct a dynamic social network $G =\left \{ G^{t_{1}}, G^{t_{2}},..., G^{t_{N}}\right \}$ at a specific time interval $\Delta_{t}$. We then acquire the node representation using the information encoding procedure established in the state-of-the-art techniques from recent study~\cite{feng2022heterogeneity}. A fully-connected layer is used to transform $\bm{x}_i^{t_{k}}$, the feature of user $i$ at timestamp $t_{k}$, into the initial user vector $\bm{h}_i^{t_{k},\{0\}}$: 
\begin{equation}
    \bm{h}_i^{t_{k},\{0\}} = \sigma (\bm{W}_I\bm{x}_i^{t_{k}}+\bm{b}_I),
    \label{eq:input}
\end{equation}
where $\bm{W}_I$, $\bm{b}_I$ are trainable parameters and $\sigma(\cdot)$ is a nonlinear activation function.

\subsection{Modeling Topological Structure}
\label{sec:method:topo}
Upon generating a dynamic social network and initial user vectors, we propose a structural module that leverages a message-passing mechanism~\cite{gilmer2017neural} to effectively model the topological structure of each snapshot. This structural module takes a snapshot $G^{t_{k}}=(V^{t_{k}}, E^{t_{k}})$ and a set of initial user vector $\left \{\bm{h}_{i}^{t_{k}}, {\forall} v_{i} \in V^{t_{k}} \right \}$ at timestamp $t_{k}$ as input. The output includes a new set of temporary node representations $\left \{\bm{s}_{i}^{t_{k}}, {\forall} v_{i} \in V^{t_{k}} \right \}$, which captures the structural information of the snapshot at $t_{k}$.

Graph attention networks ~\cite{velickovic2017graph} have shown superior performance when tackling graph data, by specifying different weights to different nodes within a neighborhood. Inspired by the Transformer architecture~\cite{vaswani2017attention}, we adopt a scaled dot-product attention mechanism for provisioning each node with the ability to learn the importance of its neighbors within a particular snapshot. Specifically, for a given a node pair $\left ( v_{i}, v_{j}  \right )$ in snapshot $G^{t_{k}}$, the attention weight can be calculated as such:
\begin{equation}
    \begin{aligned}
    \bm{q}^{t_{k}, \left \{ l \right \}}_{i} &= \bm{W}^{\left \{ l \right \}}_{q} \cdot \bm{h}^{t_{k},\left \{ l \right \}}_{i} + \bm{b}^{\left \{ l \right \}}_{q}, \\
    \bm{k}^{t_{k},\left \{ l \right \}}_{j} &= \bm{W}^{\left \{ l \right \}}_{k} \cdot \bm{h}^{t_{k},\left \{ l \right \}}_{j} + \bm{b}^{\left \{ l \right \}}_{k}, \\
    \alpha _{ij}^{t_{k},\left \{ l \right \}} &= \frac{\left \langle \bm{q}_{i}^{t_{k},\left \{ l \right \}}, \bm{k}_{j}^{t_{k},\left \{ l \right \}}   \right \rangle }{ { {\textstyle \sum_{u\in 
    \mathcal{N}_{\left (  i\right )}^{t_{k}}   }^{}} \left \langle \bm{q}_{i}^{t_{k},\left \{ l \right \}}, \bm{k}_{u}^{t_{k},\left \{ l \right \}}   \right \rangle} }, 
    \end{aligned}
    \label{eq:structural_attention_weight}
\end{equation}
where $\left \{ l \right \}$ denotes the $l$-th layer of structural module and $\mathcal{N}_{\left (  i\right )}^{t_{k}  }$ denotes the neighborhood of node $v_{i}$ at the timestamp $t_{k}$. The 
initial user vector $\bm{h}^{t_{k},\left \{ l \right \}}_{i}$ and $\bm{h}^{t_{k},\left \{ l \right \}}_{j}$ are transformed into a query vector $\bm{q}^{t_{k},\left \{ l \right \}}_{i}$ and a key vector $\bm{k}^{t_{k},\left \{ l \right \}}_{j}$. The attention weight $\alpha _{ij}^{t_{k},\left \{ l \right \}}$, which indicates the contribution of node $v_{j}$ to node $v_{i}$ at the snapshot $G^{t_{k}}$, is calculated by the exponential scale dot product function $\left \langle q, k \right \rangle = $ exp$\left ( \frac{\bm{q}\bm{k}^{T} }{\sqrt{d}}  \right )$, where $d$ is the input embedding dimension.

After obtaining the attention weight, we transform the initial user vector into a value vector and aggregate information from the neighboring nodes of $v_{i}$. A multi-head attention mechanism is employed to capture diverse patterns and dependencies in the topological structure of the social network:
\begin{equation}
    \begin{aligned}
    \bm{v}^{t_{k},\left \{ l \right \}}_{j} &= \bm{W}^{\left \{ l \right \}}_{v} \cdot \bm{h}^{t_{k},\left \{ l \right \}}_{j} + \bm{b}^{\left \{ l \right \}}_{v}, \\
    \bm{h}_{i}^{t_{k},\left \{ l+1 \right \} } &= \mathop{\Vert}\limits_{c=1}^{C}  \sigma \left (
    \textstyle \sum_{j\in \mathcal{N}_{\left (  i\right )}^{t_{k}}   }^{}\left ( \alpha _{c, ij}^{t_{k},\left \{ l \right \}} . \bm{v}^{t_{k},\left \{ l \right \}}_{c, j} \right )
    \right ),
    \end{aligned}
    \label{eq:structural_aggregation}
\end{equation}
where $\Vert$ represents the concatenation operation, $\alpha _{c, ij}^{t_{k},\left \{ l \right \}}$ denotes the attention weight computed by the $c$-th
attention head, and $\bm{v}^{t_{k},\left \{ l \right \}}_{c, j}$ denotes the corresponding value vector.

It is worth noting that we stack $L$ layers to allow nodes to capture more distant and global dependencies in the topological structure. The output of the last layer in the structural module is denoted as $\bm{s}$.

\subsection{Acquiring Temporal Dynamicity}

%%Social networks are inherently dynamic, because users are continuously joining, leaving, and interacting over time. While the structural module of BotDGT  effectively captures topological information of static snapshots, it lacks the ability to model the evolving patterns of dynamic social networks and the actions, strategies, or behavior patterns that could be changed by automated bots. 
While BotDGT's structural module can capture topological information from static snapshots, it insufficiently leverages historical context and fails to discover evolving behavior patterns of social bots. Inspired by~\cite{sankar2020dysat,ying2021transformers}, we devise a self-attention based temporal module to further make use of the temporal characteristics of social networks for bot detection. The temporal module takes as inputs a sequence of temporary representations of node $v_{i}$ at each timestamp, denoted as $\left \{\bm{s}_{i}^{t_{1}}, \bm{s}_{i}^{t_{2}},..., \bm{s}_{i}^{t_{N}} \right \}$. The module outputs a new sequence of user representations $\left \{\bm{\hat{z}}_{i}^{t_{1}}, \bm{\hat{z}}_{i}^{t_{2}},..., \bm{\hat{z}}_{i}^{t_{N}} \right \}$, where $\bm{\hat{z}}_{i}^{t_{k}}$ denotes the final representation that contains both topological and temporal feature of node $v_{i}$ at $t_{k}$. 

\subsubsection{Position Embedding Layer}
\label{sec:method:temp:position}

Since the self-attention mechanism is unaware of the nodes’ ordering information, we introduce a position embedding layer to accommodate temporal information in the sequence that can effectively reflect the dynamic nature of social networks. We consider two categories of position embedding -- absolute temporal position embedding and evolving temporal position embedding.

First, we embed the absolute temporal position~\cite{gehring2017convolutional} of each snapshot as a basis to capture ordering information as follows:
\begin{equation}
    \bm{p}^{t_{k}, AT} = E_{AT}(t_{k}),
    \label{eq:absolute_temproal_pe}
\end{equation}
where $\bm{p}^{t_{k}, AT}$ denotes the \textbf{A}bsolute \textbf{T}emporal position embedding for the timestamp $t_{k}$ and $E_{AT}$ denotes the trainable absolute temporal position embedding parameter. Note that the absolute temporal position embedding only relies on the order of the snapshot, indicating that the nodes in the same snapshot have the same absolute temporal position embedding, i.e., the absolute temporal position embedding is independent of the nodes' features.

Second, we embed two crucial temporal signals: the local clustering coefficient and bidirectional links ratio. These signals have demonstrated their utility in countering the disguised social bots~\cite{yang2013empirical} and could reveal the evolving behavior patterns over time.

$\bullet$ \textit{Local Clustering Coefficient (LCC)}: it measures the degree to which a node's neighbors are interconnected. Genuine users typically engage with acquaintances (e.g., friends, family members, and colleagues) who have similar connections and thus form closely-knit communities. By contrast, social bots are usually associated with randomly selected neighbors who lack close connectivity, which results in reduced clustering coefficients when compared with legitimate users. The position embedding of the local clustering coefficient is calculated as follows:
\begin{equation}
\resizebox{.95\linewidth}{!}{$
    \begin{aligned}
        LCC(v_{i}^{t_{k}}) = \frac{2 * |e_{v_{i}}^{t_{k}}|}{k_{v_{i}}^{t_{k}} * (k_{v_{i}}^{t_{k}} - 1)} ,\quad
        p_{i}^{t_{k}, LCC} = E_{LCC}(LCC(v_{i}^{t_{k}})),
    \end{aligned}
    \label{eq:local_clustering_coefficient_pe}
$}
\end{equation}
where $|e_{v_{i}}^{t_{k}}|$ is the number of edges between neighbors of node $v_{i}$ at the timestamp $t_{k}$, $k_{v_{i}}^{t_{k}}$ is the sum of the indegree and outdegree of node $v_{i}$ at $t_{k}$.

% \textbf{L}ocal \textbf{C}lustering \textbf{C}oefficient measures the extent to which a node's neighbors are connected to each other. Genuine users prefer to interact with acquaintances (e.g., friends, family members, and colleagues) who tend to connect with each other, forming close-knit communities. Nevertheless, social bots tend to have randomly selected neighbors who don't know each other, which results in lower clustering coefficients compared to legitimate accounts. The position embedding of the local clustering coefficient is calculated as follows:

$\bullet$ \textit{Bidirectional Links Ratio (BLR)}: it is a metric in social network analysis to assess the reciprocity between an account and its followings~\cite{yang2013empirical}. A bidirectional link appears when two accounts mutually follow each other. This metric proves particularly useful in distinguishing between genuine users, who often own higher bidirectional link counts due to reciprocal following acquaintances with mutual follow-backs, and social bots, who exhibit lower bidirectional link counts due to their indiscriminate following behavior and lack of reciprocal connections. The position embedding of the bidirectional links ratio is calculated as follows:
\begin{equation}
    \begin{aligned}
        BLR(v_{i}^{t_{k}}) = \frac{N_{blinks}(v_{i}^{t_{k}})}{N_{fing}(v_{i}^{t_{k}})}, \quad
        p_{i}^{t_{k}, BLR} = E_{BLR}(BLR(v_{i}^{t_{k}})),
    \end{aligned}
    \label{eq:bidirectional_links_ratio_pe}
\end{equation}
% \begin{equation}
%     BLR(v_{i}^{t_{k}}) = \frac{N_{blinks}(v_{i}^{t_{k}})}{N_{fing}(v_{i}^{t_{k}})}, \quad
%     \bm{p}_{i}^{t_{k}, BLR} = \bm{W}_{BLR}(BLR(v_{i}^{t_{k}})) + \bm{b}_{BLR}
%     \label{eq:bidirectional_links_ratio_pe}
% \end{equation}
where $N_{blinks}(v_{i}^{t_{k}})$ and $N_{fing}(v_{i}^{t_{k}})$ denote the numbers of bidirectional links and following interactions.

In summary, the integration of these two categories of position embeddings enables the temporal module to capture essential temporal insights from ordering information and the evolving behavior patterns of social bots.

\subsubsection{Temporal Attention Layer}
The temporal attention layer starts from gathering the outputs of the structural module and the position embedding layer:
\begin{equation}
        \hat{\bm{s}}_{i}^{t_{k}} = \bm{\bm{s}}_{i}^{t_{k}} + \bm{p}_{i}^{t_{k}, AT} + \bm{p}_{i}^{t_{k}, LCC} + \bm{p}_{i}^{t_{k}, BLR}.
    \label{eq:temproal_input}
\end{equation}
Then we pack the representations of node $v_{i}$ together across the timestamps, which is denoted as $\bm{\hat{S}}_{i} \in \mathbb{R}^{T \times F}$. Finally we perform multi-head temporal attention as follows:
\begin{equation}
    \begin{aligned}
    \bm{Q}_{i}, \bm{K}_{i}, \bm{V}_{i} &= \hat{\bm{S}}_{i}\bm{W}_{q}, \hat{\bm{S}}_{i}\bm{W}_{k}, \hat{\bm{S}}_{i}\bm{W}_{v},\\
    \bm{\hat{Z}}_{i} = \mathop{\Vert}\limits_{d=1}^{D}  softmax(&\frac{\bm{Q}_{d,i}\bm{K}_{d,i}^{T}}{\sqrt{F}} + \bm{Mask}) \cdot \bm{V}_{d,i},
    \end{aligned}
    \label{eq:temproal_transformer}
\end{equation}
where $\Vert$ represents the concatenation operation, $\bm{Q}, \bm{K}, \bm{V}$ are the queries, keys, and values transformed by trainable parameters $\bm{W}_{*} \in \mathbb{R}^{F \times F}$ respectively. $\bm{Mask} \in \mathbb{R}^{T \times T}$ is a sequence mask matrix that makes sure the node at timestamp $t_{k}$ only attends over its historical node representation. The $\bm{Mask}$ is defined as follows:
\begin{equation}
    Mask_{ab} = 
        \begin{cases}
            0 & \text{ if } a\ge  b  \\
            -\infty  & \text{otherwise}
        \end{cases}
    % \nonumber
\end{equation}

\subsection{Learning and Optimization}
The goal of BotDGT is to capture both the topological structure and the dynamic nature of social networks to classify the accounts into legitimate user and social bots. We pass the output of the temporal module into a linear layer and softmax layer for bot detection:
\begin{equation}
    \bm{\hat{y}}_{i} = softmax(
        \bm{W}_{2} \cdot (
        \sigma(\bm{W}_{1} \cdot \bm{\hat{z}}_{i} + \bm{b}_{1})
        ) + \bm{b}_{2}
    ),
    \label{eq:final_linear_softmax}
\end{equation}
where $\bm{\hat{y}}_{i}$ is the predicted output of node $v_{i}$ and $\bm{\hat{z}}_{i}$ is the representation of node $v_{i}$ obtained by temporal module.
% , and $\bm{W}_{1}, \bm{b}_{1}, \bm{W}_{2}, \bm{b}_{2}$ are trainable parameters.
Finally, we define the objective function that utilizes a binary cross-entropy function to classify node $v$ into legitimate users and social bots at each snapshot:
\begin{equation}
    Loss = \sum_{k=1}^{N} \sum_{v_{i}\in V^{t_{k}}}^{}  \left [ \bm{y}_{i}log(\bm{\hat{y}}_{i} ) +(1-\bm{y}_{i})log(1-\bm{\hat{y}}_{i} ) \right ],
    \label{eq:loss}
\end{equation}
where $N$ is the number of the snapshots, $\bm{y}_{i}$ is the ground truth label of node $v_{i}$.

% \vspace{-1em}
\section{Experiments}
In this section, we conduct extensive experiments on two benchmark datasets to answer the following questions:

\begin{itemize}[leftmargin=*]
\item \textbf{RQ1:} How does our framework perform in bot detection compared to baseline methods?
\item \textbf{RQ2:} What is the impact of removing individual architectural components on the framework's performance?
\item \textbf{RQ3:} What is the significance of capturing the dynamic nature of social networks for social bot detection?
% \item \textbf{RQ4:} How does BotDGT perform in comparison to other baselines in generating effective node representations for bot detection?
\end{itemize}
% More detailed information about the experiment
% settings and the implementation details can
% be found in the supplementary materials. Our codes will
% be made publicly available upon publication of the paper.

\subsection{Experimental Setup}
\subsubsection{Dataset}
We conduct experiments on two comprehensive social bot detection benchmark datasets (i.e., TwiBot-20~\cite{feng2021twibot} and TwiBot-22~\cite{feng2022twibot}) collected from Twitter. The datasets provide a wide range of entities and relationships, spanning the period from the inception of Twitter to the time of dataset creation, which supports our bot detection framework to model the topological structure and dynamic nature of social networks. It’s worth noting that Twibot-22 suffers from a class imbalance issue, where the number of humans is significantly larger than that of social bots.
\subsubsection{Baselines}
\renewcommand{\thefootnote}{\arabic{footnote}}
We compare BotDGT with comprehensive social bot detection methods categorized into two groups: feature-based methods and graph-based methods. Our code is publicly available on GitHub\footnotemark[1].
\footnotetext[1]{https://github.com/Peien429/BotDGT}

\textbf{Feature-based methods} generally extract the numerical features from user metadata or semantic features from textual information to identify social bots, including:

\begin{itemize}[leftmargin=*]
    \item \textbf{EvolveBot}~\cite{yang2013empirical} designs robust features that are expensive for bots to evade and utilizes machine learning classifiers to combat the evasion tactics of spammers.
    \item \textbf{Varol \textit{et al.}}~\cite{varol2017online} extracts groups of features from Twitter users and leverages random forest classifier to identify Twitter bot.
    \item \textbf{BotBuster}~\cite{ng2023botbuster} employs a mixture-of-experts approach to process user metadata and textual information, thereby improving cross-platform bot detection.
    \item \textbf{DeeProBot}~\cite{hayawi2022deeprobot} extracts features from the user account and leverages natural language processing techniques to encode textual information to learn user representations for bot detection.
    \item \textbf{SGBot}~\cite{yang2020scalable} is proposed to tackle the scalability and generalization issues in social bot detection by strategically selecting a subset of training data.
\end{itemize}

\textbf{Graph-based methods} generally interpret social networks as graphs and leverage geometric deep learning for social bot detection, including:
\begin{itemize}[leftmargin=*]
    \item \textbf{GCN}~\cite{kipf2016semi} equally aggregates features from neighbors and learns user representations, which are then passed to a linear layer for classification.
    \item \textbf{GAT}~\cite{velickovic2017graph} leverages an attention mechanism to assign diverse weights to different neighboring nodes, improving the learning of node representations.
    \item \textbf{BotRGCN}~\cite{feng2021botrgcn} is designed to construct a heterogeneous social network graph and employ relational graph convolutional networks for bot detection.
    \item \textbf{RGT}~\cite{feng2022heterogeneity} utilizes graph transformers and semantic attention to effectively model the heterogeneity of social networks for bot detection.
\end{itemize}

\begin{table*}[t]
\small
\centering
\caption{Performance of different social bot detection methods on TwiBot-20 and TwiBot-22. We run each method five times and report the average value as well as the standard deviation. The best and second-best results are highlighted with \textbf{bold} and \underline{underline}.}
\label{tab:modelCompare}
\resizebox{\textwidth}{!}{%
\begin{tabular}{@{}ll|llll|llll@{}}
\toprule
\multirow{2}{*}{Methods} &
  Dataset &
  \multicolumn{4}{c|}{TwiBot-20} &
  \multicolumn{4}{c}{TwiBot-22} \\ \cmidrule(l){2-10} 
 &
  Metrics &
  \multicolumn{1}{c}{Accuracy} &
  \multicolumn{1}{c}{F1-score} &
  \multicolumn{1}{c}{Precision} &
  \multicolumn{1}{c|}{Recall} &
  \multicolumn{1}{c}{Accuracy} &
  \multicolumn{1}{c}{F1-score} &
  \multicolumn{1}{c}{Precision} &
  \multicolumn{1}{c}{Recall} \\ \midrule

\multirow{5}{*}{feature-based} &
  EvolveBot &
  65.83$\pm$0.63 &
  69.75$\pm$0.50 &
  66.93$\pm$0.60 &
  72.81$\pm$0.41 &
  71.09$\pm$0.03 &
  14.09$\pm$0.08 &
  56.38$\pm$0.04 &
  8.04$\pm$0.05 \\

  &
  Varol \textit{et al.} &
  78.74$\pm$0.55 &
  81.08$\pm$0.48 &
  78.04$\pm$0.61 &
  84.37$\pm$0.67 &
  73.92$\pm$0.02 &
  27.54$\pm$0.26 &
  {\ul 75.74$\pm$0.31} &
  16.83$\pm$0.21 \\
  &
  BotBuster &
  78.55$\pm$0.44 &
  82.12$\pm$0.61 &
  79.85$\pm$0.74 &
  84.00$\pm$0.53 &
  74.33$\pm$0.17 &
  52.26$\pm$1.82 &
  63.32$\pm$1.47 &
  45.64$\pm$1.70 \\
  &
  DeeProBot &
  73.14$\pm$0.01 &
  77.05$\pm$0.02 &
  71.61$\pm$0.01 &
  83.50$\pm$0.04 &
  76.50$\pm$0.07 &
  24.74$\pm$0.08 &
  \textbf{80.00$\pm$0.27} &
  14.99$\pm$0.05 \\
  &

  SGBot &
  79.50$\pm$0.72 &
  84.15$\pm$0.53 &
  75.64$\pm$0.70 &
  {\ul 93.54$\pm$0.36} &
  75.53$\pm$0.25 &
  37.45$\pm$0.24 &
  74.31$\pm$0.16 &
  25.42$\pm$0.07 \\ \midrule

\multirow{4}{*}{graph-based} &
 GCN &
  83.51$\pm$0.56 &
  84.81$\pm$0.42 &
  {\ul 84.60$\pm$1.26 }&
  85.05$\pm$0.94 &
  77.83$\pm$0.96 &
  52.16$\pm$3.46 &
  71.83$\pm$1.12 &
  45.77$\pm$2.25 \\ 
  &
  GAT &
  85.04$\pm$0.38 &
  86.65$\pm$0.61 &
  83.69$\pm$1.23 &
  89.94$\pm$1.65 &
  78.65$\pm$0.19 &
  55.86$\pm$1.38 &
  71.24$\pm$0.80 &
  46.04$\pm$2.17 \\
&
  BotRGCN &
  85.83$\pm$0.38 &
  87.44$\pm$0.42 &
  83.98$\pm$0.34 &
  91.20$\pm$1.03 &
  {\ul 78.95$\pm$0.26} &
  {\ul 56.47$\pm$1.21} &
  72.38$\pm$1.42&
  {\ul 46.33$\pm$1.74}\\
&
  RGT &
  {\ul 86.53$\pm$0.47} &
  {\ul 87.74$\pm$0.62} &
  \textbf{90.37$\pm$0.64} &
  77.47$\pm$0.43  &
  77.01$\pm$0.21 &
  47.25$\pm$0.83 &
  72.80$\pm$0.76 &
  34.99$\pm$0.90 \\ \midrule

ours &
  BotDGT &

  \textbf{87.25$\pm$0.51} &
  \textbf{88.87$\pm$0.55} &
  84.44$\pm$0.39 &
  \textbf{94.24$\pm$0.37} &
  \textbf{79.33$\pm$0.22} &
  \textbf{58.15$\pm$0.74} &
  72.42$\pm$0.70 &
  \textbf{48.46$\pm$0.92} \\ \bottomrule
\end{tabular}%
}
\end{table*}

\subsection{Framework Performance (RQ1)}
We evaluate our proposed social bot detection framework along with several representative baselines on the two benchmarks and present the results in Table \ref{tab:modelCompare}. It is demonstrated that graph-based methods, which treat the social network as graphs, generally outperform feature-based methods. This could be attributed to the feature-based methods are easily circumvented by forging numerical or semantic features. The results underscore the importance of capturing topological structure of social networks for effective bot detection.

Our proposed BotDGT outperforms other static graph-based baseline models, including the state-of-the-art static graph model, in terms of accuracy, recall, and F1-score on both TwiBot-20 and Twibot-22 datasets. In comparison with the architecture of previous static graph-based methods, BotDGT not only leverages the topological structure of the social network but also incorporates a temporal module that captures the dynamic nature of the social network. The superior performance of BotDGT could be attributed to its ability to capture the historical context of social networks and model the behavior patterns of automated bots that may evolve over time to evade detection, enabling better discrimination of social bots disguised as legitimate users by interacting with other users. Further detailed analysis of the dynamicity modeling is provided in Section \ref{Temporal Module Analysis}.

%% As TwiBot-22 suffers from the class imbalance issue, where the number of legitimate users is much larger than the number of bot users, it's better to choose F1-score as the metric to evaluate the model performance. Specifically, our proposed BotDGT achieves a significant improvement of F1-score compared to the state-of-the-art bot detection approaches.

\begin{table}[t]
\centering
\caption{Results of ablation study. SM and TM denote the structural module and temporal module, respectively.}
\label{tab:ablation}
\resizebox{\columnwidth}{!}{%
\begin{tabular}{@{}l|cc|cc@{}}
\toprule
\multirow{2}{*}{Ablation Settings} & \multicolumn{2}{c|}{TwiBot-20}  & \multicolumn{2}{c}{TwiBot-22}   \\ \cmidrule(l){2-5} 
                          & Accuracy       & F1-score       & Accuracy       & F1-score   \\ \midrule
BotDGT    & 87.25$\pm$0.51 & 88.87$\pm$0.55 & 79.33$\pm$0.22 & 58.15$\pm$0.74 \\
replace SM w/ GCN             & 86.28$\pm$0.35 & 87.87$\pm$0.33 & 79.28$\pm$0.05 & 57.14$\pm$1.11 \\
replace SM w/ GAT             & 86.48$\pm$0.39 & 87.99$\pm$0.42 & 79.14$\pm$0.13 & 57.23$\pm$1.42 \\
replace SM w/ RGCN            & 86.33$\pm$0.55 & 87.92$\pm$0.73 & 79.17$\pm$0.15 & 57.70$\pm$1.76 \\
replace SM w/ RGT             & 86.22$\pm$0.20 & 87.64$\pm$0.24 & 79.28$\pm$0.05 & 57.24$\pm$1.30 \\
w/o TM                        & 85.72$\pm$0.21 & 86.99$\pm$0.39 & 78.64$\pm$0.12 &55.23 $\pm$1.17 \\
w/o $\bm{p}^{AT}$ in TM             & 86.42$\pm$0.40 & 88.13$\pm$0.54 & 79.23$\pm$0.12 & 56.51$\pm$0.54 \\
w/o $\bm{p}^{LCC}$ in TM                  & 86.25$\pm$0.51 & 87.92$\pm$0.53 & 79.25$\pm$0.21 & 56.77$\pm$0.93 \\
w/o $\bm{p}^{BLR}$ in TM                   & 86.56$\pm$0.22 & 88.08$\pm$0.24 & 79.29$\pm$0.15 & 56.84$\pm$0.98 \\ 
% replace TM w/ LSTM                       & 86.53$\pm$0.30 & 88.13$\pm$0.27 & 79.15$\pm$0.07 & 56.61$\pm$0.52 \\
% replace TM w/ GRU                        & 86.42$\pm$0.29 & 88.06$\pm$0.24 & 79.13$\pm$0.04 & 56.01$\pm$0.67 \\ 
 \bottomrule
\end{tabular}%
}
% \vspace{-1.0em}
\end{table}

\subsection{Ablation Study (RQ2)}
BotDGT comprises a structural module and a temporal module, to generate node representations for social bot detection. In this section, we conduct an ablation study on BotDGT by removing or replacing one specific component at a time to assess its significance. The components validated in this section are the message-passing mechanism in the structural module, the temporal module and the temporal position embeddings within it. Experimental results of these ablation models on TwiBot-20 and TwiBot-22 are shown in Table \ref{tab:ablation}. 

\textbf{Effect of Structural Attention.} In the structural module, we propose a scaled dot-product attention mechanism that assigns diverse weights to different neighboring nodes for propagating node messages and capturing topological structure in each static snapshot. We replace the structural attention with other static graph-based methods proposed to detect social bots, including GCN, GAT, BotRGCN, and RGT. The observed performance degradation indicates that the scaled dot-product structural attention better captures the underlying topological structure for dynamicity modeling of social network.

\begin{table*}[t]
\centering
% \vspace{-0.4em}
\caption{Performance comparison between the original static graph-based baselines and the enhanced models with the proposed temporal module. \textbf{Bold} indicates the improved model performance.}
\label{tab:multi-static-transformer}
% \vspace{-0.4em}
\resizebox{0.9\textwidth}{!}{%
\begin{tabular}{@{}ll|ll|ll|ll|ll@{}}
\toprule
\multirow{2}{*}{Dataset} &
  Metric &
  \multicolumn{2}{c|}{Accuracy} &
  \multicolumn{2}{c|}{F1-score} &
  \multicolumn{2}{c|}{Precision} &
  \multicolumn{2}{c}{Recall} \\ \cmidrule(l){2-10} 
 &
  Method &
  \multicolumn{1}{c}{Original} &
  \multicolumn{1}{c|}{Enhanced} &
  \multicolumn{1}{c}{Original} &
  \multicolumn{1}{c|}{Enhanced} &
  \multicolumn{1}{c}{Original} &
  \multicolumn{1}{c|}{Enhanced} &
  \multicolumn{1}{c}{Original} &
  \multicolumn{1}{c}{Enhanced} 
   \\ \midrule
\multirow{4}{*}{TwiBot-20} &
  GCN &
  83.51$\pm$0.56 &
  \textbf{86.28$\pm$0.35} &
  84.81$\pm$0.42 &
  \textbf{87.87$\pm$0.33} &
  84.60$\pm$1.26 &
  84.20$\pm$0.24 &
  85.05$\pm$0.94 &
  \textbf{91.88$\pm$0.56} \\
 &
 GAT &
  85.04$\pm$0.38 &
  \textbf{86.48$\pm$0.39} &
  86.65$\pm$0.61 &
  \textbf{87.99$\pm$0.42} &
  83.69$\pm$1.23 &
  \textbf{84.65$\pm$0.56} &
  89.94$\pm$1.65 &
  \textbf{91.61$\pm$1.25} \\
  &
  BotRGCN &
  85.83$\pm$0.38 &
  \textbf{86.33$\pm$0.55} &
  87.44$\pm$0.42 &
  \textbf{87.92$\pm$0.73} &
  83.98$\pm$0.34 &
  83.95$\pm$0.74 &
  91.20$\pm$1.03 &
  \textbf{91.77$\pm$1.02} \\
  &
  RGT &
  86.53$\pm$0.47 &
  86.22$\pm$0.20 &
  87.74$\pm$0.62 &
  87.64$\pm$0.24 &
  90.37$\pm$0.64 &
  84.00$\pm$0.26 &
  77.47$\pm$0.43 &
  \textbf{91.61$\pm$0.79} \\ \midrule
\multirow{4}{*}{TwiBot-22} &
  GCN &
  77.83$\pm$0.96 &
  \textbf{79.28$\pm$0.05} &
  52.16$\pm$3.46 &
  \textbf{57.14$\pm$1.11} &
  71.83$\pm$1.12 &
  71.10$\pm$1.44 &
  45.77$\pm$2.25 &
  \textbf{47.82$\pm$2.16} \\
 &
  GAT &
  78.65$\pm$0.19 &
  \textbf{79.14$\pm$0.13} &
  55.86$\pm$1.38 &
  \textbf{57.23$\pm$1.42} &
  71.24$\pm$0.80 &
  71.45$\pm$2.00 &
  46.04$\pm$2.17 &
  \textbf{48.77$\pm$2.87} \\
  &
  BotRGCN &
  78.95$\pm$0.26 &
  \textbf{79.17$\pm$0.15} &
  56.47$\pm$1.21 &
  \textbf{57.70$\pm$1.76} &
  72.38$\pm$1.42 &
  \textbf{73.37$\pm$1.65} &
  46.33$\pm$1.74 &
  \textbf{48.07$\pm$3.02} \\
  &
  RGT &
  77.01$\pm$0.21 &
  \textbf{79.28$\pm$0.05} &
  47.25$\pm$0.83 &
  \textbf{57.24$\pm$1.30} &
  72.80$\pm$0.76 &
  \textbf{72.99$\pm$1.90} &
  34.99$\pm$0.90 &
  \textbf{47.17$\pm$2.60} \\ \bottomrule
\end{tabular}%
}
\vspace{-0.5em}
\end{table*}

\textbf{Effect of Temporal Module.} The temporal module assumes a pivotal role in modeling the dynamicity of social networks and the evolving behavior patterns of social bots. We first assess the overall impact of the temporal module and then evaluate the effect of position embeddings within the temporal module. The variant \textbf{w/o TM} removes the temporal module of BotDGT, characterizing the social network as a static graph. As shown in Table \ref{tab:ablation}, the variant's performance experiences significant degradation when compared to the original BotDGT architecture, which indicates the importance of incorporating temporal information for effective bot detection. The temporal position embeddings are proposed to capture the temporal information in the sequence. The variant \textbf{w/o $\bm{p}^{AT}$ in TM} confirms the effectiveness of capturing the ordering information of the time sequence. The variants \textbf{w/o $\bm{p}^{AT}$ in TM} and \textbf{w/o $\bm{p}^{LCC}$ in TM} demonstrate the importance of considering evolving behavior patterns. Overall, these results confirm the crucial role of positional embeddings in effectively modeling the dynamicity of social networks. %% Furthermore, we replace the transformer-based temporal attention with state-of-the-art recurrent neural networks (i.e., LSTM, GRU). It is demonstrated that replacing the temporal attention with the recurrent neural network exhibits a degradation in model performance, which proves the effectiveness of our temporal attention design.
\subsection{The significance of dynamicity modeling (RQ3)} \label{Temporal Module Analysis}

To investigate the impact of exploiting the inherent dynamicity of social networks for bot detection, we enhance several static graph-based baselines with the proposed temporal module and compare their performance before and after this enhancement. The experimental results outlined in Table \ref{tab:multi-static-transformer} reveal that the majority of the graph-based baselines with the temporal module integrated exhibit noticeable performance improvement compared with their non-enhanced counterparts. 

Similar to BotDGT's result, all of the enhanced graph-based methods can achieve higher recall rates, indicating the significance of capturing the dynamic nature of social networks in detecting disguised social bots. Notably, the graph-based methods with the temporal module integrated experience a slight reduction in precision. We speculate that the increased sensitivity to social bots derived from the temporal module might lead to a slight increase in false positives. This observation also gives rise to a decrease in RGT's F1-score on the Twibot-20 dataset, which initially had the highest precision and the lowest recall rate. Nevertheless, there is a notable improvement in F1-score for most graph-based methods, which can reaffirm the importance of incorporating the dynamic nature of social networks in bot detection.

% \begin{figure}[!htbp]
%     \centering
%     \subfloat[GCN]{
%     \centering
%     \includegraphics[width=0.15\textwidth]{figures/gcn.pdf}
%     }
%     \subfloat[GAT]{
%     \centering
%     \includegraphics[width=0.15\textwidth]{figures/gat.pdf}
%     }
%     \subfloat[BotRGCN]{
%     \centering
%     \includegraphics[width=0.15\textwidth]{figures/rgcn.pdf}
%     }
%     \quad
%     \subfloat[RGT]{
%     \centering
%     \includegraphics[width=0.15\textwidth]{figures/rgt.pdf}
%     }
%     \subfloat[BotDGT w/o TM]{
%     \centering
%     \includegraphics[width=0.15\textwidth]{figures/transformer without temporal module.pdf}
%     }
%     \subfloat[BotDGT]{
%     \centering
%     \includegraphics[width=0.15\textwidth]{figures/transformer with temporal module.pdf}
%     }
%     \centering
%     \vspace{-0.4em}
%     \caption{Node representations visualization. \textcolor{red}{Red} represents bots, while \textcolor{blue}{blue} represents humans.}
%     \label{fig:rq5} 
% \end{figure}

% \subsection{Representation Learning (RQ4)}
% To answer RQ4, we adopt T-SNE~\cite{van2008visualizing} to visualize the learned user representations of BotDGT. For comparison, we also visualize the user representations learned by GCN, GAT, BotRGCN, RGT, and BotDGT without the temporal module. The visualization results presented in Figure \ref{fig:rq5} demonstrate that compared to other methods, BotDGT can learn more discriminative and comprehensive node representations in general, which also indicates that the information obtained from temporal context can aid in learning user representations for social bot detection.

\begin{figure}[t]
\begin{minipage}[t]{0.47\linewidth}
% \centering
\includegraphics[width=\linewidth]{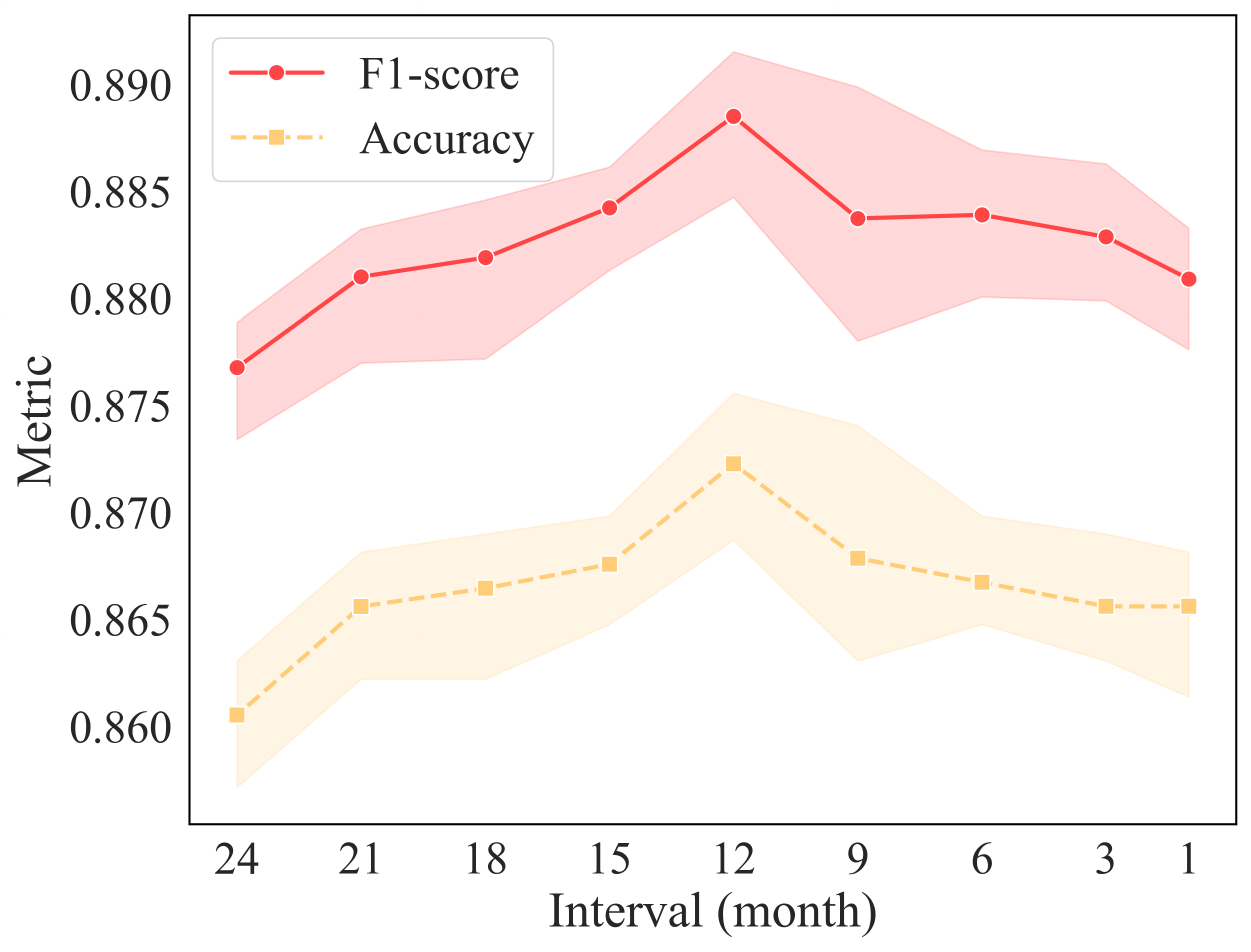}
\caption{Framework performance at various time intervals}
\label{rq3a}
\end{minipage}%
\hfill%
\begin{minipage}[t]{0.47\linewidth}
% \centering
\includegraphics[width=\linewidth]{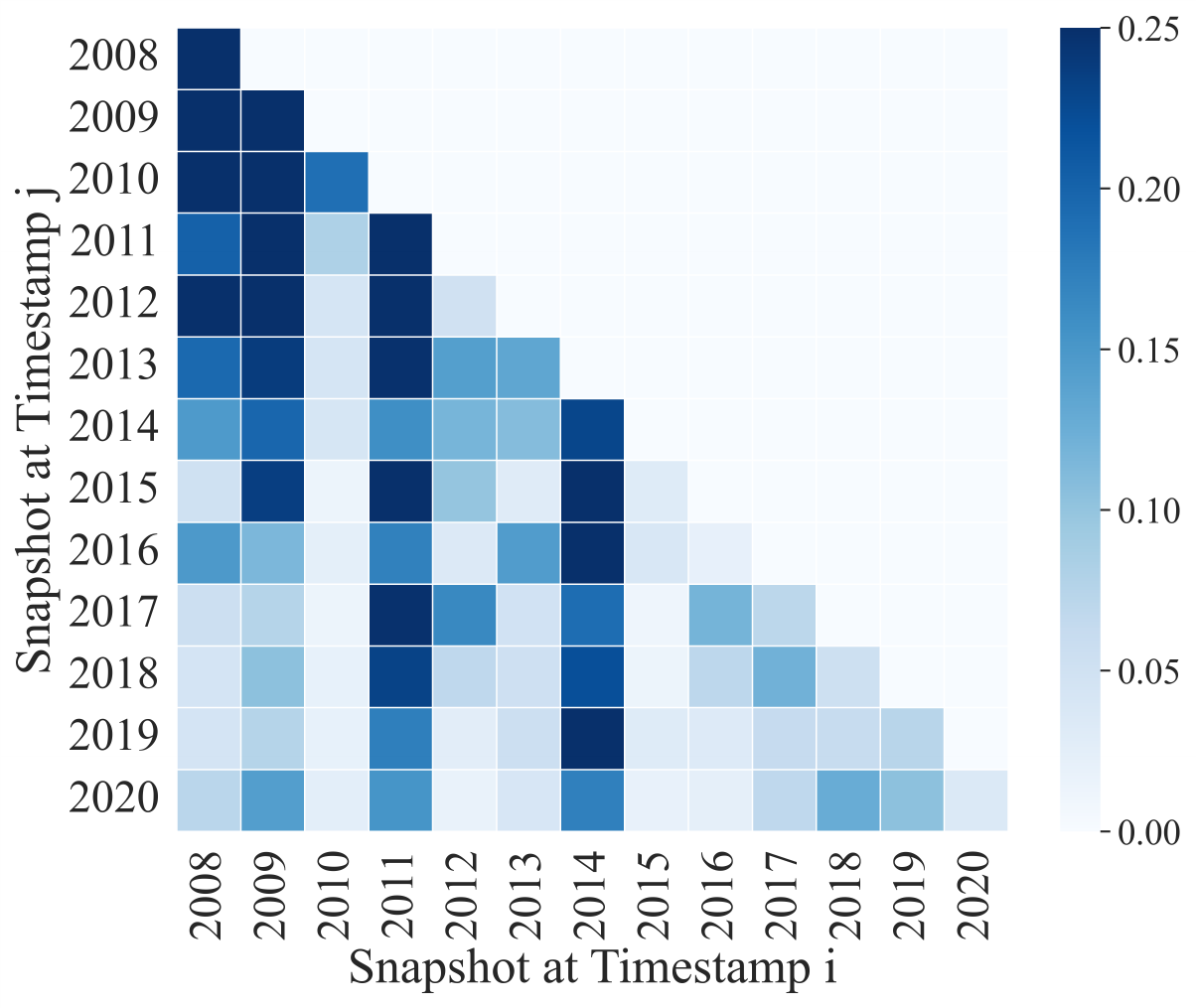}
\caption{Distribution of temporal attention weights}
\label{rq3b}
\end{minipage}
\vspace{-1.0em}
\end{figure}

\vspace{-0.5em}
\section{Discussion}
\subsection{Granularity for Dynamicity Modeling}
Different from the static graph-based approaches that rely on the most recent snapshot, BotDGT integrates historical context from multiple snapshots. During the process of graph construction in BotDGT, we construct a series of snapshots to depict the social network at a time interval $\Delta_{t}$, which determines the granularity of the dynamic social network. To explore how granularity affects dynamicity modeling, we evaluate BotDGT with various time intervals on TwiBot-20. The result, illustrated in Figure \ref{rq3a}, shows that the model performance initially improves as the granularity becomes finer due to richer temporal information. However, a decline in the model performance is observed when the granularity becomes finer than 12 months. We speculate that while setting the granularity finer than 12 months provides more historical context and detailed evolving patterns, it also introduces more noise and short-term fluctuations of the social network, making it more challenging for the temporal module to learn consistent patterns. Furthermore, the study~\cite{cresci2020decade} has found that the evolution of social bots is not very frequent, indicating that social bots mostly don't change their actions or strategies within a short period of time. Therefore, excessively fine granularity may not provide meaningful insights into the evolving behavior patterns exhibited by social bots.

\subsection{Temporal Attention for Dynamicity Modeling}
To further explore how temporal attention affects the performance of bot detection, we visualize the distribution of temporal attention weights averaged over test nodes on TwiBot-20 at the time interval of 12 months. In Figure \ref{rq3b}, each row represents the attention weight distribution of the snapshot at $t_{k}$ over its historical snapshots at $t_{1}, \cdots, t_{k-1}$. As shown in Figure \ref{rq3b}, BotDGT doesn't assign uniform weight to historical snapshots, indicating the different contributions of each snapshot for social bot detection. When predicting the most recent snapshot of the dynamic social network, BotDGT assigns more weight to the historical snapshots before 2015, rather than focusing on the more recent ones. We speculate that the reason is that advanced social bots evolved to change their behavior patterns and interact with legitimate users around 2015, which is consistent with \textit{the rise of a third wave of bots} from 2016 onwards as described in the study~\cite{cresci2020decade}. Thus, BotDGT is capable of adapting attention weight distributions to effectively incorporate historical context.

\subsection{Limitations and Future Work}
A primary limitation lies in the fact that, while BotDGT demonstrates significant improvements, we acknowledge the increased computational cost associated with dynamicity modeling. However, we believe that this trade-off is acceptable, given the potential adverse impact that social bots could cause. Another limitation is that our evaluation is limited to the Twitter platform due to the lack of datasets from other platforms and the generalizability of BotDGT to other platforms remains uncertain. We leave optimizing the computational efficiency of BotDGT and assessing its performance across diverse social media ecosystems as future work.

% Our future direction includes conducting further research on the interpretability of BotDGT and exploring techniques to enhance the computational speed of BotDGT on large-scale social networks, thereby improving its model efficiency.
\vspace{-0.5em}
\section{Conclusion}
The proliferation of social network bots has led to negative consequences. While state-of-the-art bot detection methods generally represent the social network as a static graph, they tend to overlook the dynamic nature of the social network. In this paper, we propose a bot detection framework, BotDGT, to exploit the inherently dynamic nature of social networks, which incorporates the historical context and models the evolution of behavior patterns. Experimental results on real-world datasets demonstrate that BotDGT outperforms other static graph-based bot detection methods. Further studies indicate the significance of exploiting the social network's dynamic nature for effective bot detection.
\section*{Acknowledgments}
% This work was supported by the National Key Research and Development Program of China (No. 2022YFB3104700, No. 2022YFB3105405, No. 2021YFC3300502), NSFC (No. 62322202), Beijing Natural Science Foundation (No. 4222030), Guangdong Basic and Applied Basic Research Foundation (No. 2023B1515120020), and Shijiazhuang Science and Technology Plan Project (No. 231130459A).

This work was supported by the National Key Research and Development Program of China through grants 2022YFB3104700, 2022YFB3105405, 2021YFC3300502, NSFC through grants 62322202, U2003206, Beijing Natural Science Foundation through grant 4222030, Guangdong Basic and Applied Basic Research Foundation through grant 2023B1515120020, Shijiazhuang Science and Technology Plan Project through grant 231130459A.

% This work was supported by the National Key Research and Development Program of China through grant 2022YFB3104700, NSFC through grant 62322202, Beijing Natural Science Foundation through grant 4222030, Guangdong Basic and Applied Basic Research Foundation through grant 2023B1515120020, Shijiazhuang Science and Technology Plan Project through grant 231130459A.

% \newpage
\bibliographystyle{named}
\bibliography{myPaper}
\end{document}